%% file: main.tex
\documentclass[conference,a4paper]{IEEEtran}
\usepackage{graphicx}               
\usepackage{amsmath,amssymb}
\usepackage{citesort}

\def\(({\left(}
\def\)){\right)}
\def\[[{\left[}
\def\]]{\right]}

\newtheorem{theorem}{Theorem}
\newtheorem{proposition}{Proposition}
\newcommand{\be}{\begin{equation}}
\newcommand{\ee}{\end{equation}}
\newcommand{\bea}{\begin{eqnarray}}
\newcommand{\eea}{\end{eqnarray}}

\newcommand{\defined}{\triangleq}

\begin{document}

\sloppy

%% Paper Title
%% You can use linebreaks \\ within to get better formatting as
%% desired.
\title{Variational Free Energies for Compressed
  Sensing} \author{\IEEEauthorblockN{Florent Krzakala, Andre Manoel
    and Eric W. Tramel} \IEEEauthorblockA{
    Laboratoire de Physique Statistique, \'Ecole Normale Sup\'erieure\\
    and Universit\'e Pierre et Marie Curie, Rue Lhomond Paris 75005  France\\
    ESPCI and CNRS UMR 7083, 10 rue Vauquelin, Paris 75005 France}
  \and \IEEEauthorblockN{Lenka
    Zdeborov\'a} \IEEEauthorblockA{Institut de Physique Th\'eorique\\
    CEA Saclay and URA 2306, CNRS\\ 91191 Gif-sur-Yvette,
    France.}}

%% To balance the two columns, you should reduce the text-height of
%% the last page using the following command:
%%%%%%%%%%%%%%%%%%%%%%%%%%%%%%%%%%%%%%%%%%%%%%%%%%%%%%%%%%%%%%%%%%%%%
%\addtolength{\textheight}{-9.35cm}
%%%%%%%%%%%%%%%%%%%%%%%%%%%%%%%%%%%%%%%%%%%%%%%%%%%%%%%%%%%%%%%%%%%%%
%% with an appropriate value. This command must be place on the second
%% last page, i.e., for a one-page abstract here, for a two-page
%% abstract right after the \maketitle command.

%% Create the title:
\maketitle

\begin{abstract}
  We consider the variational free energy approach for compressed
  sensing. We first show that the na\"ive mean field approach performs
  remarkably well when coupled with a noise learning procedure. We also
  notice that it leads to the same equations as those used for iterative
  thresholding. We then discuss the Bethe free energy and how it
  corresponds to the fixed points of the approximate message passing
  algorithm. In both cases, we test numerically the direct
  optimization of the free energies as a converging sparse-estimation
  algorithm. 
\end{abstract}

\section{Introduction}
\label{sec:intro}
\input{intro}

%--- Section edited. (e.w.t.)

\section{The mean field approach}
\label{sec:mean_field}
\input{mean_field}

%--- Section edited. (e.w.t.)

\section{The Bethe approach}
\label{sec:bethe}
\input{bethe}

%--- Section edited. (e.w.t.)

\section{Derivation of the Bethe Free Energy}
\label{sec:bethe_gamp}
\input{bethe_gamp}
%--- Section edited. (e.w.t.)

\section{Conclusion}
\label{sec:perspectives}
\input{perspectives}

%--- Section edited. (e.w.t.)

\section*{Acknowledgment}
This work has been supported in part by the ERC under the European
Union’s 7th Framework Programme Grant Agreement 307087-SPARCS, by the Grant DySpaN of
‘‘Triangle de la Physique,’’ and by FAPESP under grant
13/01213-8.

%% References:
%% We recommend the usage of BibTeX:
%%
\bibliographystyle{IEEEtran}
\bibliography{refs}

\end{document}

%% file: intro.tex
The last few years have witnessed spectacular advances in the
application of message passing strategies to sparse estimation and
compressed sensing (CS)
\cite{DonohoMaleki10,Rangan10b,KrzakalaPRX2012}. However, these
belief propagation (BP) based strategies often possess poor
convergence properties in many real applications. It is therefore
interesting to look for alternative approaches with similar
performance but better convergence properties. A standard
alternative is the direct optimization of the so-called Bethe free
energy \cite{yedidia2003understanding,chandrasekaran2011counting}. The
goal of the present contribution is to discuss the Bethe free energy
in the context of CS, and its relation with the iterative
thresholding and the variational mean field approach.

In compressed sensing, we wish to estimate an unknown $N-$dimensional
``sparse'' signal, $\bf x$, which has only a small ratio, $\rho$, of
non-zero elements, given the knowledge of a set of measurements of
$\bf x$, the vector $\bf y$, obtained through a linear transformation
followed by a component-wise output map. First we concentrate on
the case, ${\bf y}=F{\bf x}+\xi$, where $\xi \sim
\mathcal{N}(0,\Delta_0 \mathbf{I}_M)$ is an \textit{iid} white Gaussian noise
and $F$ is the $M \times N$ measurement matrix. The case of a more
general output $P_{\rm out}(\bf y|F{\bf x})$ will be treated in Sec.~\ref{sec:bethe_gamp}. Graphical-models
\cite{wainwright2008graphical,montanari2012graphical} are a natural
tool to use when discussing such problems in a probabilistic setting.
Here we shall assume (although it is not strictly necessary,
as shown in \cite{KrzakalaPRX2012}) the knowledge of the empirical
distribution of $\bf x$,
\be
P_0({\bf x}) = \prod_i \left[ \rho {{\cal N}(0,1)} + (1-\rho)
\delta(x_i) \right],
\label{prior}
\ee
which leads to the posterior distribution
\begin{align}
	&P({\bf x} |F, {\bf y})
	= \frac{P_0({\bf x})  P({\bf y} |F, {\bf x}) }{Z({\bf y},F)},\\
	&= \frac{1}{Z({\bf y},F)}\prod_{i=1}^N P_0(
	  x_i)\prod_{\mu=1}^M \frac{e^{-\frac{(y_\mu - \sum_{i=1}^N F_{\mu
	        i}x_i)^2}{2\Delta}}}{\sqrt{2\pi \Delta}}.
	\label{p_bayes}
\end{align}
Our goal is to perform probabilistic inference and estimate the
posterior distribution by minimizing the Gibbs free energy $\mathcal{F}$  over
a trial distribution, $P_{\rm var}$, with
\be
\mathcal{F}(\{{P_{\rm var}\}}) =  D_{KL}(P_{\rm var}||P_0({\bf x})) -\left< \log P({\bf y}|{\bf x})
\right>_{\{{P_{\rm var}\}}} \, ,
\ee
where $\left<~\cdot~\right>_{\{{P_{\rm var}\}}}$ denotes the average over
distribution $P_{\rm var}$ and $D_{KL}$ is the Kullback-Leibler divergence.

%%-------- Section Break --------%%
\subsection{Outline and Main Results}
We first discuss in Sec.~\ref{sec:mean_field} the na\"ive
mean field approach to the problem. It turns out that this approach provides remarkably
good results if one couples it with estimation of the noise
variance, $\Delta$. We find that
noise estimation for the na\"ive mean field, which was first
considered in \cite{dremeau2012boltzmann},
is indeed crucial to the performance of the na\"ive mean field for CS.
We discuss minimization of the mean field free energy as an
alternative algorithm. We also
show, perhaps surprisingly, that the mean field approach leads to the
{\it same equations} as those utilized for iterative thresholding
and demonstrate how the two approaches are, in fact,
formally related within a Bayesian framework.
%--- edited. (e.w.t.)

We then consider in Sec.~\ref{sec:bethe} the Bethe free energy
and show how it corresponds to the fixed point of the
approximate message passing (AMP)
\cite{DonohoMaleki10,Rangan10b,KrzakalaPRX2012} algorithm.
We will show through an explicit minimization that the direct optimization
of this free energy is a promising alternative approach to AMP.
Interestingly, there is a very close relationship between the
minimization of the mean field and the Bethe free energy.

Finally, in Sec.~\ref{sec:bethe_gamp} we derive the Bethe free
energy in the case of generic output distribution $P_{\rm out}$. In a
recent work \cite{riegler2013fixed} the authors have shown how a
fixed point of the generalized-AMP corresponds to a stationary
point of a function. Perhaps unsurprisingly, we show that this
function is the Bethe free energy itself.

%% file: mean_field.tex
%%-------- Section Break --------%%
\subsection{A Separable Ansatz}
It is instructive to first review the simplest variational solution to
the CS problem, namely, the mean field one where
$P_{\rm var}=\prod_i Q_i(x_i)$. In such a case, the minimum of the free
energy is achieved for $Q_i(x_i) \propto \exp{\left<-\log {\((P({\bf
      y}|{\bf x}) P_0({\bf x})\))} \right>_{P_{\rm var}({\bf x}_{\setminus i})}}$ where we denote ${\bf
  x}_{\setminus i}$ to be all entries of $\bf x$ which {\it are not}
$x_i$.
We thus observe that the
variational distribution is a product of the prior and a Gaussian
which defines the distribution for $x_i$,
\be
Q(x_i;R_i,\Sigma_i) \defined \frac 1{Z(R_i,\Sigma_i)} P_0(x_i)
e^{-\frac{(x_i-R_i)^2}{2\Sigma_i^2}},
\label{MFVAR}
\ee
with the normalization $Z(R,\Sigma)=\int {\rm d} x P_0(x)
e^{-\frac{(x-R)^2}{2\Sigma^2}}$. Note that since we
consider \eqref{MFVAR} at a single coefficient, we drop
$i$ from the notation. We denote the mean and the variance
of \eqref{MFVAR} by the functions $f_a$ and $f_c$, respectively,
\begin{align}
f_a(R,\Sigma) &\defined \int {\rm d}x \frac x{Z(R,\Sigma)} P_0(x)
e^{-\frac{(x-R)^2}{2\Sigma^2}}, \\
f_c(R,\Sigma) &\defined \int {\rm d}x \frac {x^2}{Z(R,\Sigma)} P_0(x)
e^{-\frac{(x-R)^2}{2\Sigma^2}} -f_a^2(R,\Sigma) .
\end{align}
The following identities will be useful in the sequel,
\begin{align}
\frac{\partial}{\partial R} \log Z(R,\Sigma)  &= \frac{f_a(R,\Sigma^2)-R}{\Sigma^2 },\\
\frac{\partial}{\partial \Sigma^2} \log Z(R,\Sigma) &= \frac{(f_a(R,\Sigma^2)-R)^2+f_c(R,\Sigma)}{2\Sigma^4}.
\end{align}
With this separable ansatz,
we can now compute the expression for the Gibbs free energy,
using the short-hand notations $a_i \defined f_a(R_i,\Sigma_i)$
and $c_i \defined  f_c(R_i,\Sigma_i)$,
at the mean field level for every pair $(R_i,\Sigma_i)$,
\begin{align}
  & {\cal F}_{MF}(\{R_i\},\{\Sigma_i\}) \defined \frac M2 \log{\((2\pi
     \Delta\))} + D_{\emph KL}(Q||P_0) \nonumber \\
  &
  + \frac{1}{2 \Delta}\sum_{\mu} \[[(y_{\mu} - \sum_i F_{\mu i}
  a_i)^2+ \sum_i  F_{\mu i}^2 c_i \]] \, ,
  \label{FNRG:MF}
\end{align}
where $i \in \left\{ 1,2,\dots, N\right\}$,
$\mu \in \left\{ 1,2, \dots, M \right\}$, and
the Kullback-Leibler divergence between the variational ansatz
and the prior is given by
\be D_{\emph KL}(Q||P_0) = - \sum_i \left[ \log{{Z}(R_i,\Sigma_i)} +
\frac {c_i+(a_i-R_i)^2}{2\Sigma^2_i} \right]. \label{FNRG:DL}
\ee
%
%%-------- Section Break --------%%
\subsection{Stationary Points and Iterative Thresholding}
We now investigate the stationary points of \eqref{FNRG:MF}. In order to do
so, we shall first consider a slightly different free energy that we
shall call the ``unconstrained'' mean field free energy wherein we
%% AM called -> call
treat the $a_i$ and $c_i$ in \eqref{FNRG:MF} and \eqref{FNRG:DL} as
free variables independent from $R_i$ and $\Sigma_i$. At the stationary
points of this unconstrained free energy we find
%--- Edited. (e.w.t.)
%
\bea
\frac 1{\Sigma_i^2} &=& \sum_{\mu} \frac{F_{\mu
    i}^2}{\Delta} \label{mf_1},  \\
R _i &=& a_i +\frac{\Sigma_i^2}{\Delta}  \sum_{\mu} F_{\mu i} (y_{\mu}-\sum_{j} F_{\mu j}a_j ) \label{mf_2}, \\
a_i &=& f_a(R_i,\Sigma_i),\label{consistency_a}\\
c_i &=& f_c(R_i,\Sigma_i). \label{consistency_c}
\eea
From \eqref{consistency_a} and \eqref{consistency_c} we see that
stationary points of the unconstrained and constrained free energy are equivalent.
These equations are, in fact, nothing more than the iterative mean field method, where one
updates the distribution \eqref{MFVAR} at each iteration.
In \cite{dremeau2012boltzmann} this method was applied,
albeit with different notations, sequentially to each element in $\bf x$
in order to minimize the free energy.
Properly rescaled, these equations lead to the following property.
\begin{proposition} The fixed points of iterative thresholding using a
  given thresholding function $\eta_\Delta$ are identical to the (properly
  rescaled) stationary points of the mean field free energy
  \eqref{MFVAR}.
\end{proposition}
%--- Edited. (e.w.t.)
\begin{IEEEproof}
If one rescales $F$ such that
$\sum_{\mu} F_{\mu i}^2=1$,
when the fixed-point equations are updated in parallel
we see
\be
\mathbf{a}^{t+1} =
\eta_{\Delta}(F^*\mathbf{z}^t+\mathbf{a}^t) \text{~~where~~} \mathbf{z}^t=\mathbf{y}-F\mathbf{a}^t,
\ee
where $\eta_\Delta (x) = f_a(x,\Delta)$. This is {\it exactly} iterative thresholding (see \cite{maleki2010optimally}).\end{IEEEproof}
%--- Edited. (e.w.t.)
%b
This is an interesting and, perhaps, unexpected connection which was
also noticed in the context of AMP \cite{montanari2012graphical}.
If one performs a mean field variational Bayesian learning with an $\ell_1$ or
$\ell_0$ type ``prior'', then the resulting update equations are
nothing more than soft and hard iterative thresholding, respectively.
%--- Edited. (e.w.t.)

%%-------- Section Break --------%%
\subsection{Numerical Investigation}
In order to study the performance of the mean field approach, we have
performed a numerical optimization of the mean field free energy, as shown in Fig. \ref{fig123},
using the knowledge of both the prior distribution
and the value of the true noise variance, $\Delta_0$.
Surprisingly, the results
are rather poor. Since the free energy is not
convex, it may possess many minima. Because of this,
the correct solution is almost never found in
any setting we tested.
%--- Edited. (e.w.t.)

Motivated by the results of \cite{dremeau2012boltzmann} and by
the strong connection between the AMP fixed points and noise estimation,
which we discuss in the sequel, we thus consider $\Delta$ as a further
variable to optimize over rather than a parameter. This modification
is also favorable because it allows for the inference of the noise variance,
a value which is generally unknown {\it a priori}.
The estimate of the noise variance
is given by the zero of the partial derivative of \eqref{FNRG:MF}
w.r.t. $\Delta$,
\begin{equation}
	\Delta^{*} = \frac{1}{M} ||\mathbf{y} - F \mathbf{a}||_2^2
			     + \frac{1}{M} ||{F}^2 \mathbf{c}||_1,
  \label{eq:delta_fp}
\end{equation}
which shows that $\Delta^*$ is a function of the proximity
of the means, $\mathbf{a}$, to the measurements in the projected domain and
the estimation of the variances, $\mathbf{c}$, where the square in the
second term is taken element-wise.

As shown in Fig. \ref{fig123}, when the noise variance is learned
the performance of the mean field approach improves dramatically
and displays a much better phase transition in reconstruction performance
than convex optimization (which does not use the prior knowledge of $P_0(x)$).
This transition is very close to the one obtained by AMP when the signal
is very sparse ($\rho$ small).
As noted in \cite{dremeau2012boltzmann}, the
sequential update of \eqref{mf_1}-\eqref{consistency_c} is guaranteed to
converged to a local minima of the mean field free energy.
Our goal in the next section is to have similar guarantee while matching AMP performance.
%--- Edited. (e.w.t.)

%% file: bethe.tex
AMP has been shown to be a very powerful algorithm
for CS signal recovery.
The algorithm is obtained by a Gaussian approximation of
the BP algorithm when the measurement matrix $F$ has
\textit{iid} elements of mean and variance of $O(1/N)$. We refer the reader to
\cite{DonohoMaleki10,Rangan10b,KrzakalaPRX2012} and in
particular to \cite{KrzakalaMezard12} for the present notation and the
derivation of AMP from BP. Here, we give the iterative form of
the algorithm:
\bea
V^{t+1}_\mu &=& \sum_i F_{\mu i}^{2} c_i^t \, ,\label{eq:V} \\
\omega^{t+1}_\mu  &=& \sum_i F_{\mu i} a^t_i -
(y_\mu-\omega^t_\mu)\frac{V^{t+1}_\mu}{\Delta+V^{t}_\mu} \, , \label{eq:omega}\\
(\Sigma^{t+1}_i)^2 &=& \left[ \sum_\mu \frac{F^2_{\mu
      i}}{\Delta+V^{t+1}_\mu} \right]^{-1} \, ,  \label{eq:S}\\
R^{t+1}_i &=& a^t_i + (\Sigma^{t+1}_i)^2 \sum_\mu F_{\mu i}
\frac{(y_\mu - \omega^{t+1}_\mu)}{\Delta+V^{t+1}_\mu} \, ,\label{eq:R}
\eea
together with the consistency equations \eqref{consistency_a} and \eqref{consistency_c}.
%--- Edited. (e.w.t.)

%%-------- Section Break --------%%
\subsection{AMP vs Mean Field}
\begin{figure*}[ht]
\hspace{-0.2cm}
\includegraphics[width=2.59in]{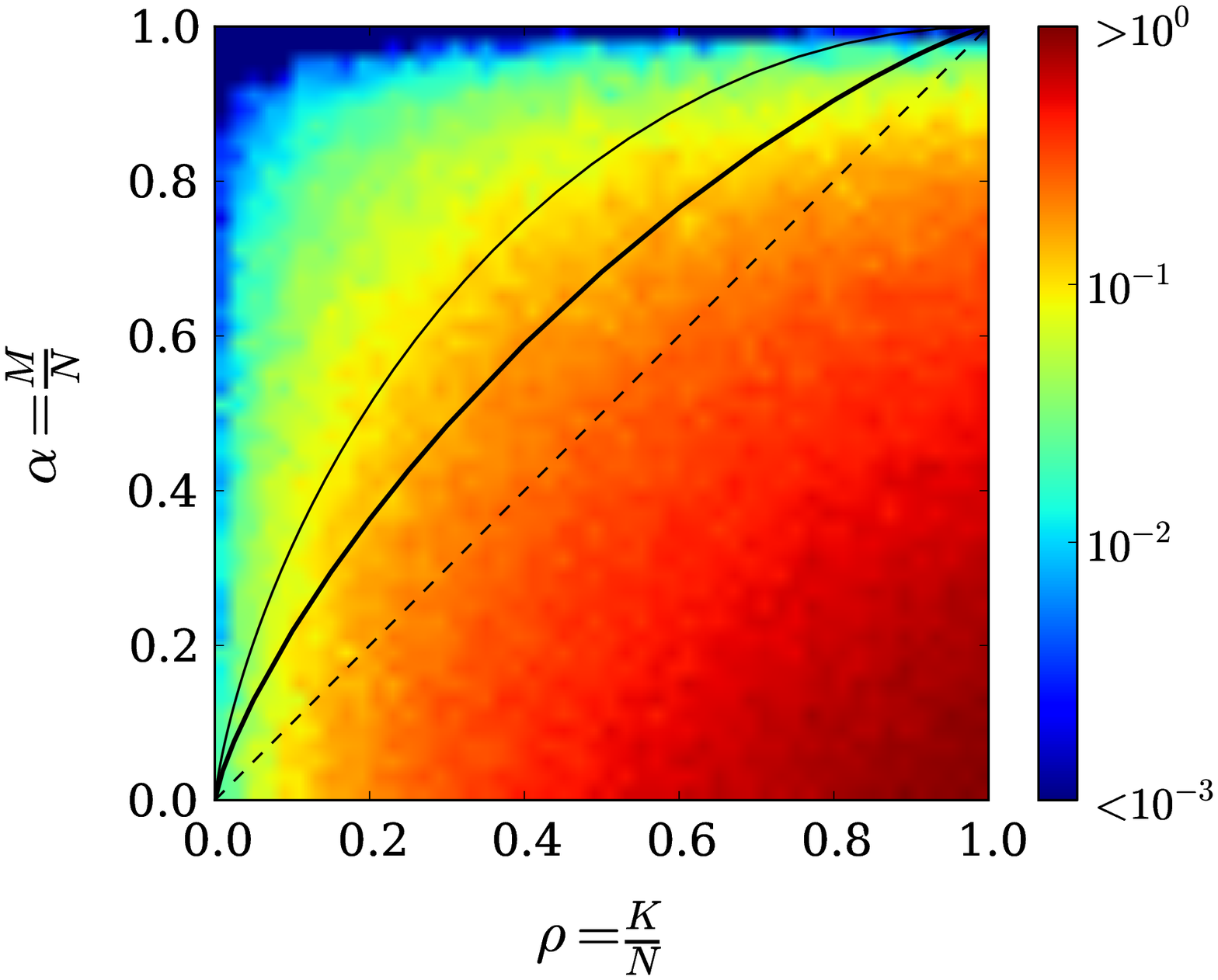}
\includegraphics[width=2.5in]{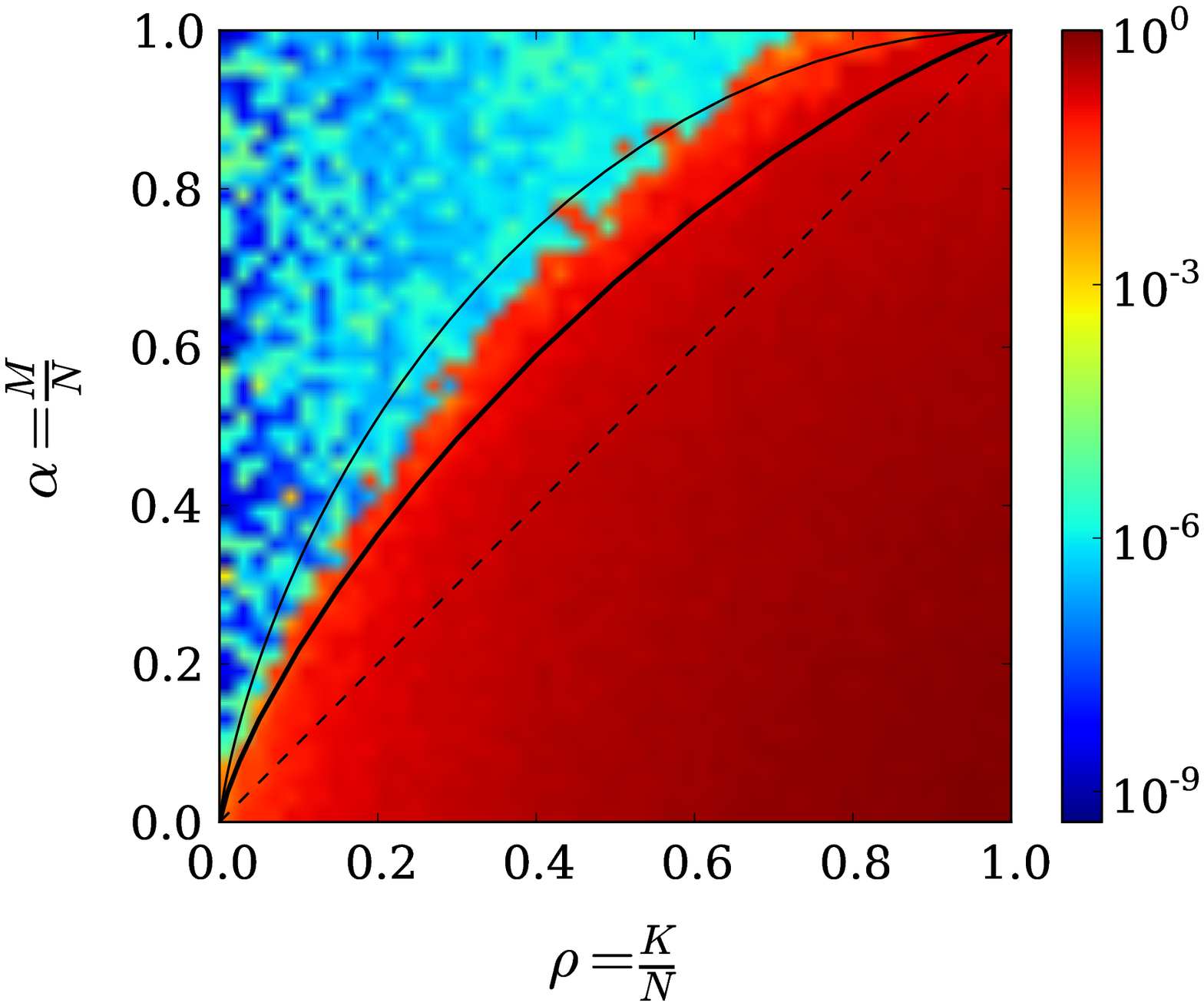}
\includegraphics[width=2.5in]{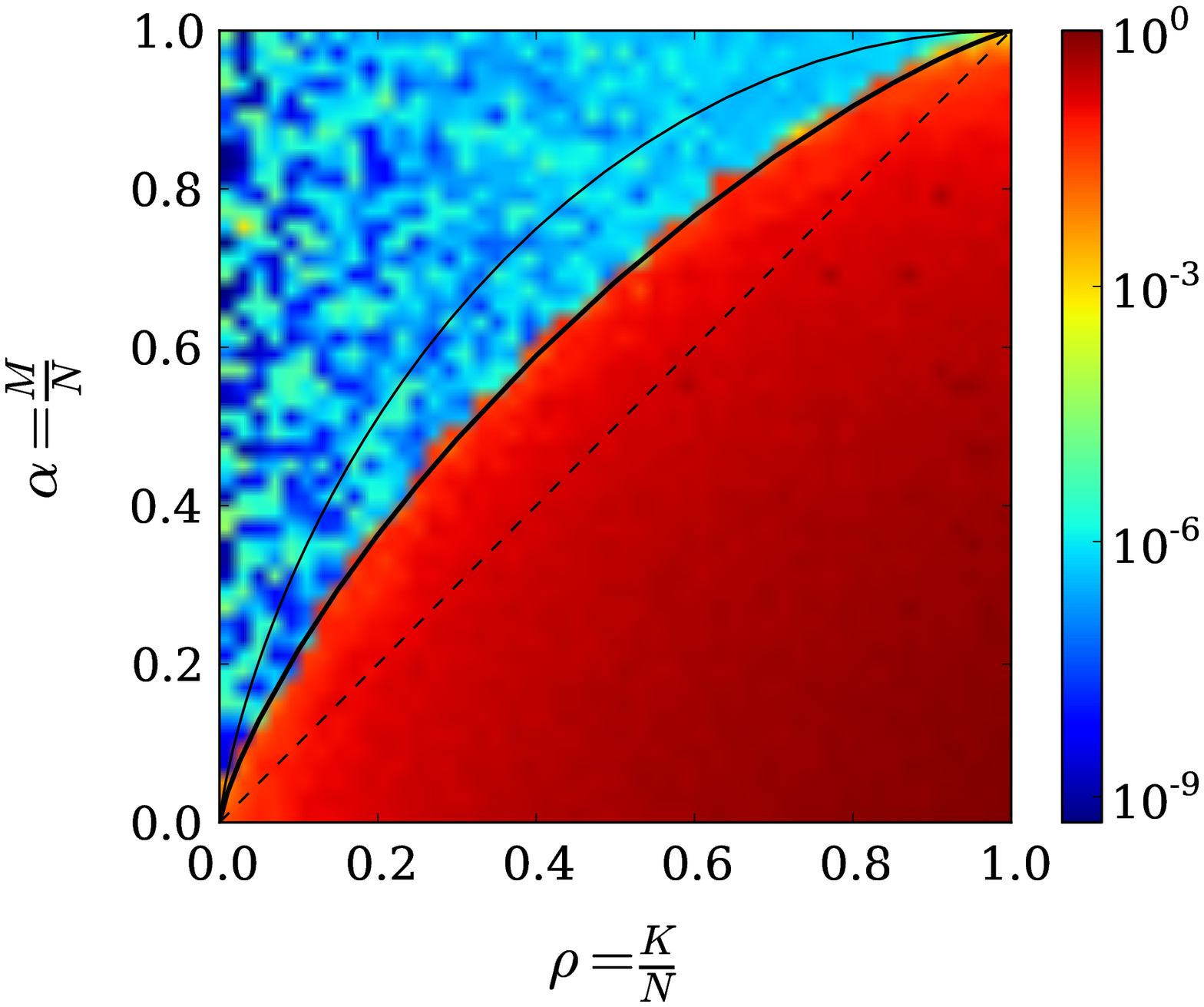}
\caption{Phase diagram for the performance of (from left to right)
  mean field, mean field with noise learning, and Bethe in the
  $\alpha$--$\rho$ plane using Gaussian noise with
  $\Delta_0=10^{-8}$, $N=1024$, $\alpha=M/N$, and $\rho=K/N$. The
  measurements ${\bf y}$ are generated using matrix $F$ with \textit{iid} Gaussian
  elements of zero mean and unit variance.
  These numerical results were obtained using the
  function \texttt{fmin\textunderscore l\textunderscore bfgs\textunderscore b}
  of the Scipy package to minimize the free energy \eqref{FNRG:MF}
  (left and center) and \eqref{FNRG:BETHE} (right).  The lines
  denote (from bottom to top) the optimal threshold for noiseless compressed
  sensing (straight dashed line, e.g. \cite{WuVerdu11}), the Bayesian AMP
  phase transition for a Gauss-Bernoulli signal (reached in the right
  panel, see \cite{KrzakalaPRX2012}) and the Donoho-Tanner transition
  for convex $\ell_1$ reconstruction \cite{Donoho05072005}.
  \textbf{Left:}
  In the pure mean field case, reconstruction is always
  mediocre.
  \textbf{Center:}
  With noise learning, the performance greatly improves and, in
  %% AM removed 'becomes'
  particular, outperforms convex optimization.
  \textbf{Right:}
  The best results
  are obtained by the minimization of the Bethe free energy which gives
  the same results as the AMP algorithm.}
\label{fig123}
\end{figure*}

We now investigate the fixed-points of AMP. In this case, one can
solve for $\omega$ in \eqref{eq:omega} and remove this
variable from all the other equations. Then, at the fixed points, we obtain
\bea
\frac 1{\Sigma_i^2} &=& \sum_{\mu} \frac{F_{\mu i}^2}{\Delta + \sum_j
  F_{\mu j}^2 c_j},  \label{b_1}\\
R _i &=& a_i +\frac{\Sigma_i^2}{\Delta}  \sum_{\mu} F_{\mu i} (y_{\mu}-\sum_{j} F_{\mu j}a_j ). \label{b_2}
\eea
These are exactly the same equations as the mean-field ones {\it
except} for ${\Sigma_i^2}$, where the $\Delta$ term has been replaced
by $\Delta + \sum_{ j}F_{\mu j}^2c_j$.
This difference is crucial to the performance of AMP over the
mean field.
The key in AMP is that the variance-like term
$\Sigma^2_i$ is computed consistently with the present estimations as it
incorporates the effect of all $c_i$. This, {\it a posteriori},
admits the interpretation of noise learning in the
mean field approach as a method of approximating \eqref{b_1} by using
$\Delta^*$ in \eqref{mf_1}. Hence, the similarity in performance between
AMP and the mean field approach with noise learning seen in Fig.~\ref{fig123} for
$\rho$ small.
%--- Edited. (e.w.t.)
%
%--- Edited. (e.w.t.)

%%-------- Section Break --------%%
\subsection{The Bethe Free Energy}
While AMP is a powerful method,
it does not always converge to a solution,
especially if the entries of the sensing matrix are not \textit{iid} randomly
distributed.  A simple modification of the mean field free
energy \eqref{FNRG:MF} leads to what is called the Bethe free
energy \cite{yedidia2003understanding,MezardMontanari09},
%--- Edited. (e.w.t.)
%
% \bea
% &&N{\cal F}_{Bethe}(R_i,\Sigma_i) \defined \sum_{\mu}
%  \frac{ (y_{\mu} - \sum_i F_{\mu i}
%   a_i)^2} {2 \Delta_{\mu}} \label{FNRG:BETHE} \\
% &+&\!\!\!\! \frac 12\log {\[[1+ \sum_i  F_{\mu i}^2 c_i/\Delta_{\mu}\]]}
% + \frac 12 \log{2\pi \Delta_{\mu}}
% + D_{\emph KL}(Q||P_0). \notag
% \eea
%\begin{align}
 % N{\cal F}_{Bethe}(R_i,\Sigma_i) \defined &\sum_{\mu}
 % \frac{ (y_{\mu} - \sum_i F_{\mu i} a_i)^2} {2 \Delta_{\mu}} \label{FNRG:BETHE} \\
 % + &\sum_{\mu} \frac 12\log {\[[1+ \sum_i  F_{\mu i}^2 c_i/\Delta_{\mu}\]]} \notag\\
 % + &\sum_{\mu} \frac 12 \log{2\pi \Delta_{\mu} }
 % + D_{\emph KL}(Q||P_0). \notag
%\end{align}
%
 \bea
 & {\cal F}^{\rm Bethe}(\{R_i\},\{\Sigma_i\}) \defined \sum_{\mu}
  \frac{ (y_{\mu} - \sum_i F_{\mu i} a_i)^2} {2 \Delta} +
  \frac M2 \log{2\pi \Delta} \notag \\
  &+\sum_{\mu} \frac 12\log {\[[1+ \sum_i  F_{\mu i}^2 c_i/\Delta\]]}
  + D_{\emph KL}(Q||P_0), \label{FNRG:BETHE}
 \eea
where the KL distance is given by \eqref{FNRG:DL}.
This free energy is derived in Sec.~\ref{sec:bethe_gamp}.
For now, let us accept this expression and investigate its properties.
\begin{proposition}\label{prop:bounded}
 The Bethe free energy ${\cal
    F}^{\rm Bethe}(R_i,\Sigma_i) $ in~\eqref{FNRG:BETHE} has at least
    one minimum and is strictly bounded by the mean field free energy.
\end{proposition}
\begin{IEEEproof}
  The proof follows from the fact that \eqref{FNRG:BETHE} is the sum
  of two terms bounded by below: a ``cost-like'' term bounded by
  $\frac {M}{2} \log{2\pi \Delta} $ and the
  non-negative Kullback-Leibler term. Moreover, since $\log(1+x) \le
  x$ for $x \ge 0$, one can see that $\frac {M}{2} \log{2\pi
    \Delta} \le {\cal F}^{\rm Bethe}(\{R_i\},\{\Sigma_i\}) \le {\cal
    F}_{MF}(\{R_i\},\{\Sigma_i\})$. \end{IEEEproof}

We shall now connect this minimum, and the other possible stationary points, to the fixed point(s) of
the AMP recursion.
%--- Edited. (e.w.t.)
%
%%-------- Section Break --------%%
\subsection{Equivalence with AMP}
\begin{theorem} All stationary points of the Bethe free energy correspond to
  fixed points of AMP.
\end{theorem}
\begin{IEEEproof}
  The proof follows the same outline as the mean field free energy.
  Define the ``unconstrained'' Bethe free energy where $a_i$ and
  $c_i$ are free variables. Stationarity with respect to $c_i$ and
  $a_i$ leads to \eqref{b_1} and \eqref{b_2}, and stationarity
  w.r.t. $R_i$ and $\Sigma_i$ to the consistency equations,
  \eqref{consistency_a} and \eqref{consistency_c}.
  This demonstrates the correspondence between AMP fixed points and
  the stationary points of the ``unconstrained'' free energy.
  Since, at the stationarity points, the consistency equations are satisfied, then all
  stationarity points of the ``unconstrained'' free energy are
  stationarity points of the normal Bethe free energy
  \eqref{FNRG:BETHE}  and vice-versa.
\end{IEEEproof}
Note the difference between the ``unconstrained'' and the
``constrained'' Bethe free energy \eqref{FNRG:BETHE}. While the former
allows one to easily generate the AMP fixed points (a classical
property, see \cite{MezardMontanari09}), it is not bounded and cannot
generically be interpreted as a variational free energy. Only the
later ``constrained'' form should be considered a proper variational
functional, as indicated by Proposition \ref{prop:bounded}. Indeed,
while we should look for a minimum of the constrained free energy,
all stationary points of the unconstrained functional appear, instead, as
saddles. In fact, for a given variable $i$, the sign of the second
derivative shows that the unconstrained free energy is a minimum for
$a_i$ and a maximum for $c_i$, $R_i$ and $\Sigma_i^2$. This is, again,
reminiscent of the known phenomena that the fixed points of the Bethe free
energy are, in general, only saddles
unless some consistency conditions are imposed
(see \cite{yedidia2005constructing}).
%--- Edited. (e.w.t.)

%%-------- Section Break --------%%
\subsection{Numerical Investigation}
There are a number of ways the Bethe free energy \eqref{FNRG:BETHE}
can be used. For instance, one can utilize it to damp,
self-consistently, the AMP iteration to ensure a strict minimization,
or at least a minimizing trend, at each AMP step. We have
empirically observed that this method significantly increases the
convergence properties of the AMP
approach\footnote{See
          our recent implementation of the AMP algorithm at
          \text{http://aspics.krzakala.org/} or on GitHub at
          \text{https://github.com/jeanbarbier/BPCS\textunderscore
          common}}.
Some authors \cite{Rangan10b,parker2013bilinear} have, instead, used
the mean field free energy to the same effect.
This approach, however, does not seem well justified as it is truly the
Bethe free energy which is optimized by AMP.
%--- Edited. (e.w.t.)

We have also performed numerical optimization of the Bethe free
energy using the same approach as we used for the mean field approach.
As shown in Fig. \ref{fig123}, direct minimization gives the same
performance as iterating the AMP equations and, in fact, reaches
the usual AMP limit obtained by the (rigorous
\cite{BayatiMontanari10}) state evolution
analysis \cite{KrzakalaPRX2012,KrzakalaMezard12}.
Direct minimization of the Bethe free energy is therefore a promising
alternative to the AMP when convergence problems are encountered.
%--- Edited. (e.w.t.)

%% file: bethe_gamp.tex
Given the probabilistic model defined in Sec.~\ref{sec:intro} and that the
measurement matrix $F$ posses \textit{iid} elements of mean and variance $O(1/N)$,
the fixed point of the BP equations can be used to estimate the posterior likelihood.
The logarithm of this normalization is, up to a sign, called the Bethe free
energy \cite{yedidia2003understanding,MezardMontanari09}.
In BP, one utilizes a graphical model by updating messages from constraints to
variables, $m_{\mu \to i}(x_i)$, and from variables to constraints,
$m_{i\to \mu}(x_i)$.
Following \cite{KrzakalaMezard12}, one can write the Bethe free energy as
\be -{\cal F}^{\rm Bethe}= \sum_\mu
\log{ {\cal Z}^\mu} + \sum_i \log{{\cal Z}^i} - \sum_{\mu i} \log{{\cal Z}^{\mu
    i}}\, ,\label{Bethe}
\ee
where
\bea
{\cal Z}^i&=& \int {\rm d} x_i \prod_{\mu} m_{\mu \to i}(x_i) P_0(x_i) \, ,\\
{\cal Z}^{\mu i}&=& \int {\rm d}x_i m_{\mu \to i}(x_i) m_{i \to
  \mu}(x_i)\, , \\
{\cal Z}^\mu&=& \int {\rm d} z \frac{e^{-\frac{(\omega_{\mu} -z)^2}{2V_{\mu
        }}}}{\sqrt{2\pi V_{\mu }}} P_{\rm
  out}(y_{\mu}|z) \, .  \label{eq:Zmul}
\eea
%
%--- edited. (e.w.t.)
Note that here we follow the framework of GAMP \cite{Rangan10b}
and consider the context where the observations of the sparse signal are
given by element-wise measurements, $y_{\mu }$, specified by some known
probability distribution function $P_{\rm out}(y_{\mu }|z_{\mu})$,
where $z_{\mu}=\sum_i F_{\mu i} x_i$.
%\bea P_{\rm out}(Y|Z)=
%\prod_{\mu} P^{\mu}_{\rm out}(y_{\mu }|z_{\mu }) \, .
%\label{PY}
%\eea
Following the notation of \cite{Rangan10b}, we define the
output function as \be g_{\rm out} (\omega, y , V) \defined \frac{ \int
  {\rm d}z P_{\rm out}(y|z)\, (z-\omega) \,
  e^{-\frac{(z-\omega)^2}{2V}} }{ V \int {\rm d}z P_{\rm out}(y|z)
  e^{-\frac{(z-\omega)^2}{2V}} } \, . \label{eq:def_gout} \ee
%--- edited. (e.w.t.)

The integrals in the evaluation of the free energy are not
algorithmically tractable in their general form. Using the same
notations as in~\cite{KrzakalaMezard12} and the same
approximations
used to go from BP to AMP, which are valid in the leading order when $N\to\infty$,
we shall obtain a tractable form for the
free energy. First, we use the properties of the BP messages
\cite{KrzakalaMezard12} to rewrite \eqref{Bethe},
\be -{\cal F}^{\rm
  Bethe} = \sum_{\mu} \log{{\cal Z}^{\mu}} +\sum_{i} \log{
  {\cal X}^{i}} + \sum_{i}   {\cal Y}^i,
\ee
where
\begin{align}
{\cal X}^{ i } &= \int {\rm d}x_{ i } P_0(x_{i }) e^{- \frac{x^2_{ i}
  }{2
    \Sigma^2_{i }} +  x_{i }  \frac{R_{ i }}{\Sigma^2_{ i }} } \,
, \label{eq:chi} \\
  {\cal Y}^i
&= - \frac{R_{i }}{ \Sigma^2_{ i }} a_{ i  } + \frac{1}{2 \Sigma^2_{i
    }} (c_{i } + a^2_{i })+ \frac{1}{2}  c_i\sum_{\mu=1}^M F^2_{\mu i}
  g^2_{\rm out} \, .
%{\cal X}^{ i  \to \mu } &= &\int {\rm d}x_{i } P_X(x_{ i })
%e^{-\frac{x^2_{ i }}{2N}\sum_{\nu \neq \mu } A_{\nu  \to i } +
%  \frac{ x_{i }}{\sqrt{N}} \sum_{\nu\neq \mu} B_{\nu  \to i } }. \nonumber
\end{align}
%--- edited. (e.w.t.)
%
%We now expand the last term in $1/N$, keeping the leading order,
%
%\be
%\sum_{ \mu} \log{ \frac{ {\cal X}^{ i  \to \mu }}{{\cal X}^{i}} }
%= - \frac{R_{i }}{ \Sigma_{ i }} a_{ i  } + \frac{1}{2 \Sigma_{i
%    }} (c_{i } + a^2_{i })+ \frac{1}{2N}  \sum_{\mu=1}^M V_{\mu}
%  g^2_{\rm out},
%\ee
%
Then, we replace $g^2_{\rm out}$ by its fixed-point value,
$g^2_{\rm out} = (\sum_i F_{\mu i} a_i - \omega_\mu)/V_\mu$,
to obtain the following expression which gives the (negative) posterior
likelihood given a {\it fixed point} of the GAMP equations,
\begin{align}
&{\cal F}^{\rm Bethe}_{\rm GAMP}\left(\{R_i\},\{\Sigma_i\},\{\omega_{\mu}\},
  \{a_i\},\{c_i\}\right) =
-\sum_{\mu} \log {\cal Z}_{\mu} \nonumber \\
&- \sum_i \frac {c_i+(a_i-R_i)^2}{2\Sigma^2_i} - \sum_{\mu} \frac
{(\omega_{\mu}-\sum_i F_{\mu i}
  a_i)^2}{2V_{\mu}} \nonumber  \\
&- \sum_i \log{{Z}(R_i,\Sigma_i)}\, \text{~~with~}V_{\mu} = \sum_i
F_{\mu i}^2c_i\, ,
\label{FREE_ENERGY_BETHE_GENERATING}
\end{align}
where $Z(R,\Sigma)$ is the same as in \eqref{MFVAR}. In its
present form, the Bethe free energy can be easily computed and
satisfies the following theorem.
%--- edited. (e.w.t.)
\begin{theorem}
  \emph{(Bethe/GAMP correspondence)} The fixed points of the GAMP
  message passing equations are the stationary points of the cost function
  ${\cal F}^{\rm Bethe}_{\rm GAMP}$ \eqref{FREE_ENERGY_BETHE_GENERATING}.
\end{theorem}
\begin{IEEEproof}
  By setting the derivatives of \eqref{FREE_ENERGY_BETHE_GENERATING}
  with respect to $R_i$, $\Sigma_i$, $\omega_\mu$, $a_i$, and $c_i$ to zero we obtain
  \eqref{consistency_a}, \eqref{consistency_c}, \eqref{eq:omega},
  \eqref{eq:R}, and \eqref{eq:S}, respectively. Or, more precisely, the GAMP
  analogs of the equations using the generic output function $g_{\rm out}$.
\end{IEEEproof}
%--- edited. (e.w.t.)

While the fixed points of the message passing equations are the
stationary points of
${\cal F}^{\rm Bethe}_{\rm GAMP}$, they have no reason to minimize~\eqref{FREE_ENERGY_BETHE_GENERATING}. Indeed, they are only
saddle points of this expression. This is no surprise: we are not only
optimizing the free energy with respect to a given distribution, we also have to
satisfy the consistency conditions between, for instance, the
parameters $\Sigma_i$ and $R_i$ and the values $a_i$ and $c_i$. Only when
$a_i=f_a(R_i,\Sigma_i)$ is there consistency between these variables. In fact,
as is always the case with the Bethe free energy, only at a fixed point
can it be interpreted as an estimation of the posterior. It is
thus practical to return to a variational form of the free energy
that one should simply minimize. To do this, we impose
the consistency conditions and express the free energy as a
function of the parameters of our trial distributions for the two
matrices,
\be
  {\cal F}^{\rm B}_{\rm var}
  \left(\{R_{i}\},\{\Sigma_{i}\}\right) = {\cal F}^{\rm
  Bethe}_{\rm GAMP}\left(\{R_{i}\},\{\Sigma_{i}\},\{\omega^*_{\mu}\},
  \{a^*_{i}\},\{c^*_{i}\}\right)\, , \nonumber
\ee
%--- edited. (e.w.t.)
%
where the $*$ variables are given in terms of the fixed points as
function of $R_{i}$ and $\Sigma_{i}$ only.
In order to write this variational expression in a nicer form, let us
define the distribution
\be {\cal M}(z,\omega,V) \defined \frac{1}{{\cal
    Z}^\mu} P_{\rm out}(y|z) \frac{1}{\sqrt{2\pi V}}
e^{-\frac{(z-\omega)^2}{2V}}. \label{eq:Mz}
\ee
Then, one has
\be
-D_{\rm KL}( {\cal M}|| P_{\rm out} ) = \log {\cal Z}^{\mu} +
\frac {\log 2\pi V +1+V (\partial_{\omega} g_{\rm out} +
g_{\rm out}^2)}2.
\nonumber
\ee
Finally, we obtain
\begin{align}
& {\cal F}^{\rm B}_{\rm var} \left(\{R_{i}\},\{\Sigma_{i}\}\right)
= \sum_{i} D_{\rm KL} (Q||P_0)  +\sum_{\mu} D_{\rm KL}( {\cal
    M}|| P_{\rm out} )  \nonumber \\
&+ \frac 12 \sum_{\mu}\left( \log{2\pi V_\mu^*}
+1+V_\mu^*\partial_{\omega} g_{\rm out} \right)  , \label{eq:Bethe}
\end{align}
with $V_\mu^*$ and $\omega_\mu^*$ satisfying their respective fixed-point
conditions. Note that this is in the same form as the expression in~\cite{riegler2013fixed}.
Thus, we observe that \eqref{eq:Bethe} is the Bethe estimation of
the posterior. An important difference with \cite{riegler2013fixed} is
that we use the Bethe free energy in Sec.~\ref{sec:bethe} in order to
obtain a bounded variational expression which is in turn used to recast AMP and GAMP as
cost minimization problems; however, \cite{riegler2013fixed}
discusses a promising ADMM-like strategy. Lastly, one can
observe that, in the case of CS with Gaussian noise corrupted measurements,
$g_{\rm out}(\omega_{\mu},y_{\mu},V_{\mu}) = (y_{\mu} - \omega_{\mu})/(\Delta + V_{\mu})$.
This output function can be used to obtain \eqref{FNRG:BETHE}.
%--- edited. (e.w.t.)

%% file: perspectives.tex
We have considered the variational free energy approach for
CS and discussed the properties of the
resulting mean field and Bethe functional. We also
demonstrate how the mean field approach paired with noise learning
serves as an approximation of the AMP algorithm.
Most interestingly, AMP has been recast in a form equivalent to
a cost function minimization. One possible
avenue for future work is to investigate efficient ways
of minimizing this cost function with convergence guarantees.
% Might be also interesting to solve
% directly the unconstrained one (we know who's positive and
% negative, etc....) Promising directions...